\documentclass [] {aa} 
\usepackage{epsfig}

\def\kms{\,km\,s$^{-1}$} 
\def\ms{\,m\,s$^{-1}$} 
\def\msd{\,m\,s$^{-1}$\,d$^{-1}$} 

\def\msol{$M_{\odot}$}

\begin{document}

\thesaurus{08(03.20.7;07.19.2; 08.02.4; 08.09.02 GJ86; 08.12.2)} 
\title{A planet orbiting the star Gliese\,86\thanks{Based on
observations collected at the La Silla Observatory, ESO Chile, with the echelle 
spectrograph CORALIE at the 1.2m Euler Swiss telescope}}

\author{D.  Queloz\inst{1}$^,$\inst{2} \and M.  Mayor\inst{1}   \and L.  
Weber\inst{1} \and A.  
Bl\'echa\inst{1} \and M.  Burnet\inst{1}  \and B.  Confino\inst{3} \and D.  
Naef\inst{1} \and F. Pepe\inst{1}
\and N. Santos\inst{1} \and S.  Udry\inst{1}}
\institute{Observatoire de Gen\`eve, 51 ch.  des
	   Maillettes, CH--1290 Sauverny, Switzerland\and 
	   Jet Propulsion Laboratory,
           Mail-Stop:  306-388, 4800 Oak Grove Drive,
	   Pasadena, CA 91109, USA\and 
	   St-Luc Observatory,
	   Switzerland}

\offprints{Didier.Queloz@obs.unige.ch}
\date{Received 20 May 1999/ Accepted 1 Octobre 1999} 

\maketitle

\begin{abstract} 
A 4\,M$_J$ planet with a 15.8\,day orbital period has been detected from 
very precise radial velocity  measurements  with the CORALIE echelle spectrograph. A second remote and more  massive companion has also been detected.\\ All the planetary companions so far detected in orbit closer than  0.08\,AU have  a parent star with a statistically higher metal content compared to  the metallicity distribution of other stars with  planets. Different processes occuring  during their formation may provide a possible explanation for this observation.

\end{abstract}


\section{The search for extra-solar planets with CORALIE}

Almost 20 planetary  companions with minimum masses ($m_2\,\sin\,i$) less than 
10\,$M_J$ have so  far been detected by very high-precision radial  velocity 
surveys (\cite{Butler99}, \cite{Marcy99}, \cite{Fischer99} and  \cite{MCM99} for a review of older detections).
The semi-major  axes of  their orbits  range from very small (0.05\,AU)  to 
3\,AU. Some have eccentric orbits,  others have  secondary more massive 
companions, some have both. The  large observed  spread in orbital 
characteristics of all known planetary candidates  causes some  difficulties in understanding their formation  process in comparison with  our own solar system. It also raises the issue of the real nature of  these objects, particularly the more massive ones.  
 
In June 1998 we initiated a systematic and large scale exoplanet search survey 
(1600 nearby G and K  stars) in the southern hemisphere with  the new  1.2\,m 
alt-azimuth Euler Swiss telescope  at  La Silla, ESO Chile. The technique we are using to detect planets is to  look for a stellar reflex motion due to  an orbiting planet by very precise radial velocity measurements. The CORALIE echelle 
spectrograph is used to measure star spectra from which  the Doppler effect is  
then computed. 

CORALIE  is an improved version of the ELODIE spectrograph (\cite{Baranne96}) 
with which, 4 years ago, the first extra-solar planet  orbiting a star (51\,Peg)
was discovered (\cite{Mayor95}). The  CORALIE front-end  adaptator is located at
the Nasmyth focus of the Euler telescope. Two sets of two fibers can   
alternatively feed the spectrograph which is located in an isolated and 
temperature  controled  room. The set of fibers used for high precision radial 
velocity measurements includes a   double scrambler device designed by Dominique Kohler (see \cite{Queloz99} for references) to 
improve the stability of the input  illumination of the spectrograph.
Thanks to  a slightly different optical combination at the entrance of the 
spectrograph and  the use of a 2k by 2k CCD camera with smaller pixels 
(15$\mu$m),  CORALIE has a larger resolution than ELODIE. A resolving power of 
50,000 ($\lambda/\Delta\lambda$) is observed with a 3 pixel sampling. As with 
the ELODIE spectrograph, CORALIE makes use of on-line reduction software 
that computes the radial velocity of stars several minutes after 
their observation. (See \cite{Baranne96} for details about the reduction 
process). The simultaneous thorium technique is used to correct any 
instrumental drifts occuring during the star exposure (see  \cite{Queloz99} for details). The many improvements carried out in the thermal 
control and the resolution of the instrument, as well as in the reduction 
software, yield   a factor two improvement in the instrument precision compared
with ELODIE.

\section{A planet orbiting Gliese\,86}

Gliese\,86  (HD13445, HIC 10138) is a bright ($m_V= 6.12$) early K dwarf 
($B-V=0.81$, $T_{\mbox{eff}}=5350\,K$,   $\log(g)=4.6$, \cite{Flynn97}) from the
southern hemisphere, in the Eridanus  (River) constellation. It is a  close 
star, 10.9\,pc  away from our Sun ($\pi=91.6$\,mas, 
measured by the Hipparcos satellite). Its absolute  magnitude is 6.257, yielding (with $BC=-0.2$)  a luminosity $L=0.27\,L_\odot$. It is a high proper motion  star, slightly  metal poor ([Fe/H]$=-0.24$, \cite{Flynn97}). It has low chromospheric activity ($\log{R'}_{\textrm{HK}}=-4.74$,  \cite{Saar97a}). No rotational broadening has been detected (\cite{Saar97a}) and there is only an upper limit  on the Li content in its atmosphere (N(Li)$<-0.24$,  \cite{Favata97}). From Hipparcos photometry, the star is stable ($\sigma(\textrm{H}_p)=0.008$). In summary, Gliese\,86 bears all the characteristics of a few billion year old K 
dwarf from the old disk population.  In the H-R diagram, Gliese\,86 
lies  slightly below the ZAMS.  However we believe that there are enough  
uncertainties in the temperature and bolometric 
correction estimates of Gliese\,86  -- stemming  from its low metal content -- 
to   believe that its location in the HR diagram below the ZAMS is not 
significant.

A 15.8\,day period radial velocity variation has been detected from CORALIE 
measurements (Fig.\,1). In Table~1 are listed the orbital elements of the best 
fit solution (least square) for an orbital motion after correction of a 0.36\msd
linear drift (see below).  Assuming a 0.8\msol~ for the primary and that the 
radial velocity effect is caused by the orbital motion  of the star, we  
conclude that  a 4\,M$_J$  companion  (minimum mass) is orbiting Gliese\,86.

\begin{table}
\caption[]{Orbital elements of Gliese 86 after correction of the 0.36\msd linear 
drift of
 the $\gamma$-point. }
\begin{tabular}{lrll}
\hline
\noalign{\smallskip}
$P$             & 15.78     &$\pm$    0.04    & d\\
$T$             & 2451146.7 &$\pm$    0.2     &   d\\
$e$             & 0.046     &$\pm$    0.004   &\\
$V_r^\dagger$   & 56.57     &$\pm$    0.01    &\kms\\
$\omega$             & 270       &$\pm$    4       & $\degr$\\ 
$K_1$           & 380       &$\pm$    1       & \ms \\
$f_1(m)$        & $8.9\cdot10^{-8}$ &$\pm$  $0.1\cdot10^{-8}$& $M_{\sun}$\\ 
$(O-C)^\ddagger$&7          &                 &\ms \\    
$N$             & 61        &                 &\\
\noalign{\smallskip}
\hline
\end{tabular}
\\  ($^\dagger$) At $T_0=2451150$\,d
\\  ($^\ddagger$) Without the drift correction the O-C of the fit would be 
13\ms
\end{table}

\begin{figure}
\psfig{width=\hsize,file=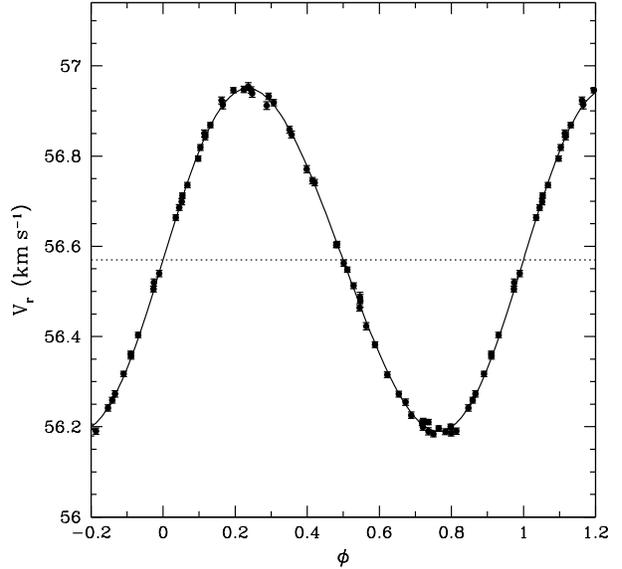}
\caption[]{Phased orbital motion of Gliese 86 corrected from the long term 
drift. The solid line is the best fit orbit. See orbital  elements in Table 1}
\end{figure}

The  planetary companion to Gliese\,86 is close to its host star  with a 
0.11\,AU  semi-major orbital  axis. It has a low, although $99$\% significant non-zero, eccentricity (\cite{Lucy71}). The 7\ms residual from the fit indicates very low intrinsic intrumental errors from night to  night, taking into account that each measurement has  approximately   5\ms photon noise error and could be 
affected as well by some low level radial velocity variations intrinsic to the stellar  atmosphere. Such low instrumental error agrees with the  instrumental error   measured by  $P(\chi^2)$ analysis of all the stars of our sample so far observed (about 300). See \cite{Duq91} for a detailed description of the instrument error estimate by the $P(\chi^2)$ statistic.

A long term drift of the radial velocity  (0.5 \msd ) is observed  from 20
years of CORAVEL measurements  (Fig.~2).  With the 300\ms typical precision of  
CORAVEL radial velocities, the short orbit  is marginally detected in the last 
measurements. Interestingly, with the recent  CORALIE  measurements a smaller 
0.36\msd drift is observed (Fig.~3). A statistical analysis of the reliability of the drift correction shows that an orbital solution without drift correction has 0.0001\% chance to occur ($\chi^2\approx 210$). The probability jumps to 40\% when the linear  drift correction is taken into account. A conservative 7\ms instrumental error is assumed for this calculation.

The period measurement of the short period planetary companion is still not accurate enough to correct the  old CORAVEL data from their extra scattering and obtain a precise drift estimate from these measurements. Thus, the  difference in drift slope between the old CORAVEL measurements and the recent CORALIE measurements is perhaps significant, but remains to be confirmed by further measurements during the course of the next season.

The long term radial velocity  variation is the signature of a remote and more 
massive  companion. The use of  historical radial velocity data together with  
the CORAVEL and the CORALIE observed  drifts  suggest a stellar companion with 
a  period longer than 100\,yr (semi-major axis larger than 20\,AU). A direct 
detection would be worth attempting since the star is close to us.

\begin{figure}
\psfig{height=\hsize,,angle=270,file=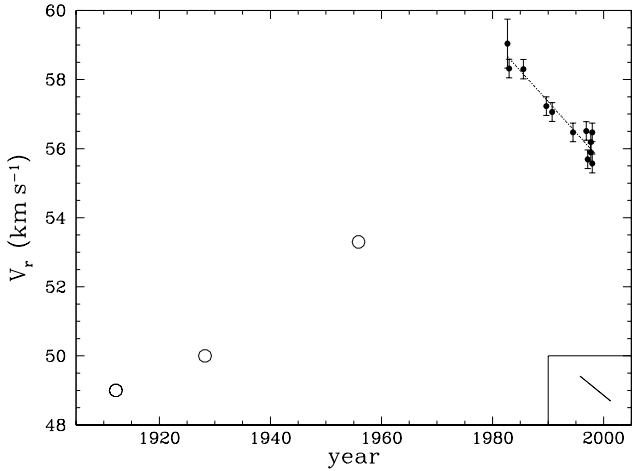}
\caption[]{Filled dots: radial velocity drift observed with CORAVEL.  A 
mean 0.5\msd  variation of the radial velocity of Gliese\,86  is measured 
({\it dotted line}). The 15.8\,day  reflex motion from the planet is marginaly seen as an extra scattering  in the last CORAVEL measurements. In the right lower box is displayed for comparison the $\gamma$-point drift measured with CORALIE. (Note that the time scale is artificially extended for the sake of a better display). Open dots: previous measurements found in the literature. No error bars are displayed but a typical 2-5\kms ~error may be assumed for these measurements}
\end{figure}  

\begin{figure}
\psfig{width=\hsize,,file=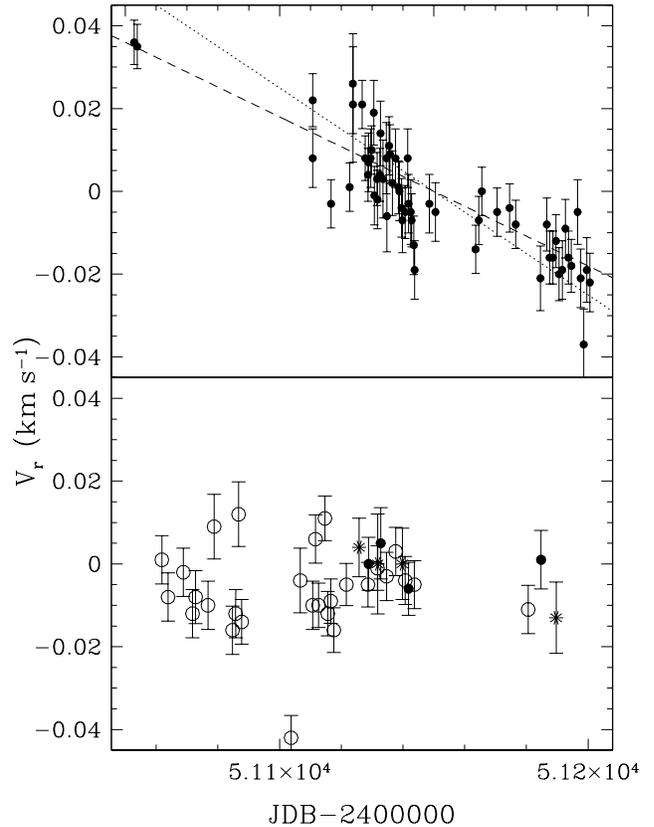}
\caption[]{{\bf Top:} CORALIE observed $\gamma$-point drift (residuals from the 
short orbit 
fit). The dashed line is the best linear fit after correction of the short 
period from the planetary companion. It corresponds to a linear drift of 0.36\msd. The dotted line is the 0.5\msd drift measured over  20\,yr CORAVEL data. {\bf Bottom:} Set of 3 non-variable stars measured at the same period of time (open dots: HD10700, stars: HD39091, filled dots: HD67199. No instrument zero point  drift is observed}
\end{figure} 

Alternative explanations to a low mass companion to explain the observed 
15.8\,day period radial velocity change  of Gliese\,86 would be activity related phenomena (\cite{Saar97b}). However Gliese\,86 doesn't exhibit any of the classical activity signatures seen on young stars, as for exemple HD166435 (Queloz et al. in prep). Gliese\,86 has no chromospheric activity. No rotational broadening is detected either, and its photometry is very stable.
Therefore, the planetary hypothesis is  most likely the correct interpretation 
for the observed  periodic radial velocity changes.

\section{Discussion}

The observed  orbital charateristics of planets are  the direct outcome of 
their formation processes and of their evolutions.  Therefore, these characteristics may be used to retrace their  formation mechanisms and to constrain theories of planetary formation. The recent spectroscopic studies of   stars  where planets have been detected have  shown that the host  star  itself may also bear marks from  some processes occuring  during   planetary formation (\cite{Gonz97}). More specifically a large number of planets with short-period orbits  have surprinsingly  metal rich  host stars. These  planets, very close to their stars, are usually referred as 51\,Peg like, or  ``hot  Jupiters''. The metallicity of their host star is much higher than  the ``average field star"   and is  not  the result of a selection process in the survey samples (\cite{MCM99}). Typical metallicities  similar to field stars may be assumed for the stars from various surveys, since these star samples  have not been selected from  any metallicity criteria.

If we look in more detail at all the planets with  semi-major axis less than 
1\,AU, where  the number of detections  is significant and not strongly high-mass biased, we observe a  relation  between the semi-major axes of 
the  planetary orbits and the metal content of  their  host stars. All  planets with semi-major axes less than  0.08\,AU   seem to have  a star with  an unusually very high metal content compared to other stars with planets (see Fig.\,4). Actually, a comparison of the two distribution using the Kolmogorov-Smirnov test indicates  a 99\% probability that the two distributions are indeed  different.

The unusual metal content of the  short 
orbit planets  had  been  pointed out shortly after the detection of 51\,Peg 
(\cite{Mayor95}). But now, with the large number  of detections of similar 
systems and others with slightly larger semi-major axes,  we observe a typical 
distance (or period) for  which this unusual high metal content is 
systematically observed.  A possible explanation may be related to some very specific processes  occuring during  the formation  of these  very close systems. However the large incertainties on the estimation of the
age of these systems  and the small mass range of primaries are noteworthy. Therefore it is difficult to completely rule out a stellar population effect.

\begin{figure}
\psfig{width=\hsize,,file=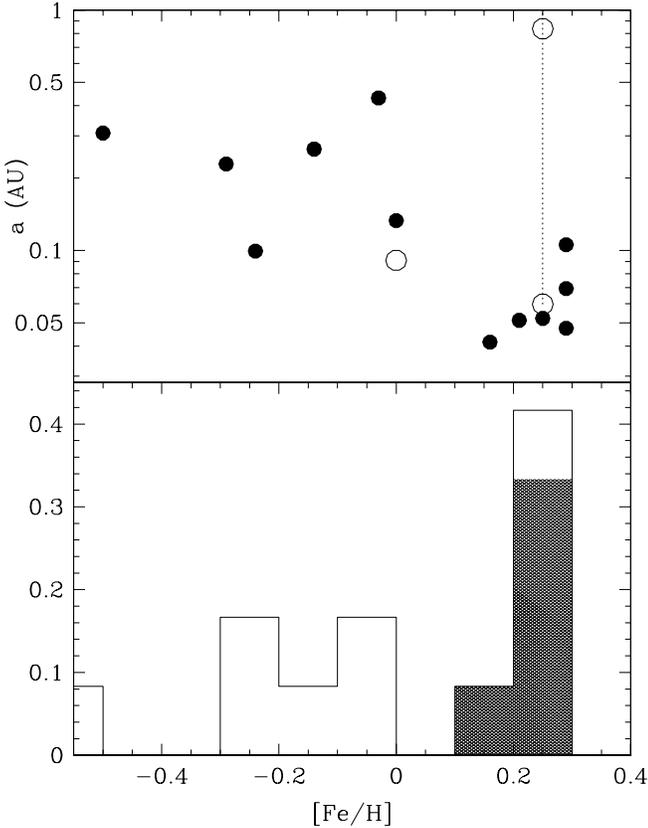}
\caption[]{{\bf Top} [Fe/H] content of  stars versus semi-major axis of the planetary orbit for  all known planetary candidates
with a $m_2\sin i <12$\,M$_J$ and $a\leq 1$\,AU. [Fe/H] measurements  are from \cite{Gonz98} and
\cite{Gonz99}. A typical 0.06 error are given by the authors. The dotted line connects the dots representatives of the two inner
planets orbiting the star  $\upsilon$\,Andromedae.\\{\bf Bottom}
Distribution of the metallicity  of stars with a planet (solid line). The
shaded area indicates stars with a planet closer than 0.08\,AU. 
Note the $\upsilon$\,And multiple system and the planet orbiting the late M star GJ876 are not included in the histogram (open dots on Top diagram)}
\end{figure}

The migration theory (\cite{Lin86}, \cite{Lin96}, \cite{Ward97}, 
\cite{Trilling98}) is one of  theories that has been  called for  to explain the
existence of very close planets that were not described by  the ``classic" 
solar-system   planetary formation model  (\cite{Boss95}, \cite{Lissauer95}). 
But so far, we have a poor  understanding of the  way the planet stops its migration. The two different  metallicity distributions pointed out in this article are perhaps a new clue to a better understanding  of the migration process or the likelyhood of an in-situ formation (\cite{Bodenheimer99}).

Others scenarios involving  strong gravity interactions with other  planets   
have been proposed as a possible origin for small planetary orbits 
(\cite{Weidenschilling96}, \cite{RasioF96}). Since these models are purely 
driven by dynamical interactions it seems  a priori difficult to expect  any   
metallicity enhancement  effect. Moreover such scenarios do not really explain
the very small orbit planets like 51\,Peg. However, if one believes that the high metal content of the star is the  end-result of  a planet swallowed by the star (\cite{Sand98}),  the gravity  interaction is a possible means to send planets into their stars.    

The  precision of surveys from which the planets have been found so far has been
limited to the  detection of systems with a $V_r$-amplitude ($K$) larger than  
25\ms. Therefore it is still premature to compare the mass of the two sets of 
planets because there is  a direct  relationship between the amplitude of the 
radial velocity curve, the semi-major axis,  and the minimum mass of the  planet
that can be detected. However if we  limit our comparison to  a  sample free of 
such bias, including  planets only having semi-major axes smaller  than 0.3\,AU,
we may be inclined to believe that  metal rich  stars tend to host on average  
less  massive planets than  solar stars or metal poor stars. 
 
The distribution of mass of all planets that have been detected
can also be studied with a  restricted sample of  planetary systems in order to 
avoid  a biased selection towards  small orbits and massive  systems. We see 
that with a sample restricted to minimum masses greater than 1\,M$_J$ and semi-major axes  smaller  than 1.3\,AU,  the number of planets per mass bin is almost  constant from 1 to 5\,M$_J$  and then drops suddenly for more massive 
companions. This reinforces the idea that  a {\sl maximum planet  mass} may lie   somewhere close to  5-7\,M$_J$ as  pointed out earlier by \cite{Mayor98}.

New discoveries and improved detection precision will allow us to get a better picture of the relation between the mass and certain orbital characteritics of planets and some peculiarities seen in the atmosphere  of their host stars. 
This will perhaps enhance our understanding of the mechanisms 
of planetary formation.

\begin{acknowledgements}
We are grateful to all the staff  who actively participated
into the building and the rapid commissioning of the new 1.2m Euler telescope 
and the CORALIE echelle spectrograph at la Silla and in particular  D. Huguenin, C. Maire, E. Ischi, G. Russinielo, M.  Fleury. 

We also thanks all the staff of the Haute-Provence observatory having
contributed to the construction of the spectrometer CORALIE, specially
D. Kohler and D. Lacroix as well as Andre Baranne (from Marseilles Observatory)
who designed the optic.

We thanks N. Molawi for his help in the 
computation of Gliese\,86 evolution tracks and P.R. Lawson for  many 
improvements  and corrections to the text.
We thanks the Geneva University and the Swiss NSF (FNRS) for their continuous 
support for this project. 
Support from FCT to N.S. in the form of a scholarship is gratefully
acknowledged. 

\end{acknowledgements}


\newpage


\begin{thebibliography}{} 

\bibitem[Baranne et al. 1996]{Baranne96}
 Baranne A., Queloz D., Mayor M., Adriansyk G., Knispel G., et al. 1996, 
 A\&A Suppl. Ser 119, 1.
 
\bibitem[Bodenheimer et al. 99]{Bodenheimer99}
Bodenheimer P., Hubickyj O. Lissauer J.J., 1999,
Protostars and Planets IV, eds. 
V. Mannings, A. Boss, S. Russell, in press

\bibitem[Boss 1995]{Boss95}
Boss A. P., 1995,
Science 267, 360

\bibitem[Butler et al. 1999]{Butler99}
Butler R.P., Marcy G.W., Fischer D. et al, 1999, 
ApJ in press.

\bibitem[Duquennoy et al. 1991]{Duq91}
Duquennoy A., Mayor M., Halbwachs J.-L., 1991, A\&AS, 88, 281.


\bibitem[Favata et al. 1997]{Favata97}
Favata F., Micela G., Sciortino S., 1997,
A\&A 322, 131.

\bibitem[Fischer et al. 1999]{Fischer99}
Fischer D., Marcy G.W., Butler R.P., Vogt S.S., Apps K.,
1999, PASP 111, 50.

\bibitem[Flynn \&  Morell 1997]{Flynn97}
Flynn C., Morell O., 1997,
MNRAS 286, 617.

\bibitem[Gonzalez 1997]{Gonz97}
Gonzalez G., 1997,
MNRAS 285, 403.

\bibitem[Gonzalez 1998]{Gonz98}
Gonzalez G., 1998,
A\&A 334, 221.


\bibitem[Gonzalez et al. 1999]{Gonz99}
Gonzalez G., Wallerstein G., Saar S.H., 1999,
ApJ 511, L111.

\bibitem[Lin \& Papaloizou 1986]{Lin86}
Lin D.N.C, Papaloizou J., 1986,
ApJ 309, 846.

\bibitem[Lin et al. 1996]{Lin96}
Lin D. N. C., Bodenheimer P., Richardson D. C., 1996,
Nature 380, 606. 

\bibitem[Lissauer 1995]{Lissauer95}
Lissauer J. J., 1995,
Icarus 114, 217

\bibitem[Lucy \& Sweeney 1971]{Lucy71}
Lucy L. B., Sweeney  M. A., 1971,
AJ 76, 544.


\bibitem[Marcy et al. 1999a]{Marcy99}    
Marcy G.W., Butler R.P., Vogt S.S., Fischer D., Liu M.C., 1999a,
ApJ 520, 239.


\bibitem[Marcy et al. 1999b]{MCM99}
Marcy G.W., Cochran W.D., Mayor M., 1999b, Protostars and Planets IV, eds. 
V. Mannings, A. Boss, S. Russell, in press

\bibitem[Mayor \& Queloz 1995]{Mayor95}
 Mayor M., Queloz D., 1995, 
 Nature 378, 355.   
 
\bibitem[Mayor et al. 98]{Mayor98}  
 Mayor M., Queloz D., Udry S., 1998,        
in "Brown Dwrafs and Extrasolar Planets", 
ASP Conf. Ser., vol. 134 

\bibitem[Queloz et al. 1999]{Queloz99}
 Queloz D.,  Casse M., Mayor M., 1999, 
 IAU Colloquium 170,``Precise stellar radial velocities", Victoria BC Canada, 
 eds J.B. Hearnshaw and C.D. Scarfe, ASP Conference Series (1999), in press.  


\bibitem[Rasio et al. 1996]{RasioF96}
Rasio F. A., Ford E. B., 1996, 
Science 274, 954.

\bibitem[Saar \& Osten 1997]{Saar97a}
Saar S. H., Osten R. A., 1997,
MNRAS 284, 803.

\bibitem[Saar \& Donahue 1997]{Saar97b}
Saar S. H., Donahue R. A., 1997,
ApJ 485, 319.

\bibitem[Sandquist et al. 1998]{Sand98}
Sandquist E., Taam R.E., Lin D.N.C., Burkert A., 1998,
ApJ 506, L65.

\bibitem[Trilling et al. 1998]{Trilling98}
Trilling D. E., Benz W., Guillot T. et al., 1998,
ApJ 500, 428.

\bibitem[Ward 1997]{Ward97}
Ward W.R., 1997
ApJ 347, 490.

\bibitem[Weidenschilling \& Marzari 1996]{Weidenschilling96}
Weidenschilling S. J., Marzari F., 1996,
Nature 384, 619.


\end{thebibliography}
\end{document}